\documentclass[12pt,preprint]{aastex}
\usepackage{emulateapj5}

\newcommand{\simgt}{\lower 2pt \hbox{$\, \buildrel {\scriptstyle >}\over {\scriptstyle\sim}\,$}}
\newcommand{\simlt}{\lower 2pt \hbox{$\, \buildrel {\scriptstyle <}\over {\scriptstyle\sim}\,$}}

\newcommand{\clover}{H~1413+117}

\newcommand{\chandra}{{\emph{Chandra}}}

\slugcomment{Received 2003 September 12; accepted 2004 January 7}
\shorttitle{Microlensing of the Cloverleaf Quasar}
\shortauthors{CHARTAS ET AL.}

\tighten

\begin{document}
\def\sarc{$^{\prime\prime}\!\!.$}
\def\arcsec{$^{\prime\prime}$}
\def\beginrefer{\section*{References}%
\begin{quotation}\mbox{}\par}
\def\refer#1\par{{\setlength{\parindent}{-\leftmargin}\indent#1\par}}
\def\endrefer{\end{quotation}}

\title{{\it CHANDRA} OBSERVATIONS OF THE  CLOVERLEAF QUASAR H~1413+117: A UNIQUE LABORATORY FOR MICROLENSING STUDIES OF A LoBAL QUASAR}

\author{
G. Chartas,\altaffilmark{1}
M. Eracleous,\altaffilmark{1}
E. Agol,\altaffilmark{2,3}
\& S. C. Gallagher\altaffilmark{4}}

\altaffiltext{1}{Astronomy and Astrophysics Department, Pennsylvania State University,
University Park, PA 16802, chartas@astro.psu.edu, mce@astro.psu.edu}

\altaffiltext{2}{Theoretical Astrophysics, Caltech, Pasadena, CA 91125, agol@tapir.caltech.edu}

\altaffiltext{3}{{\it Chandra} Fellow}


\altaffiltext{4}{University of California, Los Angeles, Division of
Astronomy \& Astrophysics, Los Angeles, CA 90095, sgall@astro.UCLA.EDU}

\begin{abstract}

We present new results uncovered by a re-analysis of a \chandra\
observation of the gravitationally lensed, low-ionization broad absorption 
line (LoBAL) quasar  \clover.  Previous analyses 
of the same \chandra\ observation led to the detection of 
a strong, redshifted Fe~K$\alpha$ line
from the combined spectrum of all images. We show that the redshifted Fe~K$\alpha$ line 
is only significant in the brighter image A. 
The X-ray flux fraction of image A is larger by a factor of 1.55 $\pm$ 0.17
than the optical R-band flux fraction, indicating that image A is significantly enhanced in 
the X-ray band.
We also find that 
the Fe~K$\alpha$ line and the continuum are
enhanced by different factors. 
A microlensing event could explain both
the energy-dependent magnification and the significant detection of Fe~K$\alpha$
line emission  in the spectrum of image A only.
In the context of this interpretation we provide constraints on the
spatial extent of the inferred scattered continuum and reprocessed Fe~K$\alpha$ line emission regions
in a LoBAL quasar. 

\end{abstract}

\section{Introduction}
Recently, the gravitationally lensed (GL) 
quasar \clover\  (the Cloverleaf quasar) was observed for $\sim$ 40~ks with the 
{\sl Chandra X-ray Observatory} (Oshima et al. 2001, Gallagher et al. 2002).
The original goal of this observation (PI: M. Hattori) was to detect a
cluster of galaxies that is thought 
to contribute to the lensing of this system (Kneib et al. 1998; Chae \& Turnshek 1999). 
X-ray emission from the cluster lens was not detected. However, 
a remarkable iron emission feature at $E$=6.21$\pm$0.16 keV (90\% confidence) 
was discovered in the combined spectrum of all lensed images.
This redshifted emission feature was interpreted as Fe~K$\alpha$ fluorescence 
from the far side of the quasar outflow.
 
\clover\  is a gravitationally lensed system
consisting of four images of a distant ($z$=2.55) broad absorption line
quasar  (Hazard et al. 1984; Magain et al. 1988).  $\sl HST$ NICMOS-2 
observations indicate that the lens is located approximately at 
the geometrical center of the four images.
A firm spectroscopic redshift of the lens has yet to be obtained, however, 
a cluster of galaxies at a redshift of $z$ = 1.7 has been reported 
to contribute to the lensing of this system (Kneib et al. 1998).
\clover\ was the first quasar to be detected in the submillimeter continuum
and in carbon monoxide emission (Barvainis et al. 1995).
\clover\ is also a low-ionization broad absorption line 
quasar (LoBAL)
showing low-ionization lines of Al III and Fe III (Monier et al. 1998).
LoBALs are BAL quasars that show one or more of the 
low-ionization lines of
Mg II, Al III, Fe II and Fe III (e.g.,  Boroson \& Meyers 1992).
Its first X-ray detection was made with the ROSAT PSPC (Chartas et al. 2000).
X-ray observations of BAL quasars indicate that they are
heavily absorbed with measured column densities ranging
between 10$^{21}$ and 10$^{24}$ cm$^{-2}$.
The extreme X-ray
weakness of LoBALs, their low values of $\alpha_{\rm OX}$,
and their high levels of UV continuum polarization 
are taken to imply that the absorbing gas in the direct line of
sight to the quasar central engine is Compton-thick to X-ray radiation from the quasar nucleus 
and only scattered or reflected radiation can reach the observer along 
an indirect line of sight (Green et al. 2001; Gallagher et al. 2002). 

Here, we present a re-analysis of the \chandra\ observation 
of \clover\  that places constraints on the 
spatial structure of the X-ray iron line and continuum scattering regions of this LoBAL quasar.
The main difference between our and previous analyses is that we 
performed a spatial and spectral analysis of the individual
X-ray images whereas previous X-ray studies were based on the combined spectrum of all images. 
Throughout this paper we adopt a cosmology in which
$H_{0}$ = 75~$h_{75}$ km~s$^{-1}$~Mpc$^{-1}$
and $q_{0}$ = 0.5.

\section{Observations and Data Analysis}
\clover\ was observed with \chandra\ for about 38~ks on 2000 April 19.
Details of  this observation were described in Oshima et al. (2000) and
Gallagher et al. (2002). 
For preparing the \chandra\ data for analysis we used the CIAO 2.2 and CALDB 2.12 products
provided by the \chandra\ X-ray Center (CXC).
To improve the spatial resolution we
removed a $\pm$ 0\sarc25 randomization applied to the event positions
in the CXC processing and employed a sub-pixel resolution technique
developed by Tsunemi et al. (2001) and later improved by Mori et al. (2001).
The deconvolved {\sl Chandra} ACIS image of H~1413+117 is shown in Figure 1.
A comparison between the deconvolved X-ray image and the optical {\it HST} images 
of \clover\ taken from the 
CfA-Arizona Space Telescope Lens Survey (CASTLES)\footnote{{\it HST} images of \clover\
in the F160W, F702W and F814W {\it HST} bands are available from the
Castles web site: http://cfa-www.harvard.edu/glensdata/Individual/H1413.html.}
clearly shows the enhancement of X-ray image A compared to that of
the optical image A.
The X-ray image positions are in agreement within errors with 
those of the {\it HST} images indicating that the
X-ray emission used in our spatial-spectral analysis originates from the lensed quasar and not
from extended emission of the lensing galaxy or host galaxy.
To account for the recently discovered decay in the quantum efficiency of ACIS
(possibly caused by molecular contamination of the ACIS filters),
we have applied a time-dependent correction 
implemented in the \verb+XSPEC+ model \verb+ACISABS1.1+.\footnote{ACISABS is an XSPEC model contributed to the \chandra\
users software exchange world wide web-site 
http://asc.harvard.edu/cgi-gen/cont-soft/soft-list.cgi.} The ACIS quantum efficiency decay is insignificant at energies above 1~keV 
and does not affect the main results of our analysis.
To estimate the X-ray flux ratios we modeled the \chandra\ images of A, B,  C and D with 
point spread functions (PSF's) generated by the simulation tool \verb+MARX+ (Wise et al 1997).
The X-ray event locations were binned with a bin-size of 0\sarc0246.
The simulated PSFs were fitted to the \chandra\ data by minimizing the
Cash $C$-statistic formed between the observed and simulated images
of \clover. The relative positions of the images were fixed to
the observed NICMOS values taken from CASTLES.
We find that the X-ray flux ratios in the full (0.2-8keV) band are 
[B/A]$_{full}$ = 0.49 $\pm$ 0.08, [C/A]$_{full}$ = 0.35$\pm$ 0.07, 
[D/A]$_{full}$ = 0.37 $\pm$ 0.07 and the number of detected X-ray events in
images A, B, C, and D were 147 $\pm$ 13, 72 $\pm$ 10, 52 $\pm$ 8, and 54 $\pm$ 9, respectively.
These flux ratios are consistent with the
values obtained from the deconvolved X-ray image shown in Figure 1.
For comparison the {\sl HST} R-band flux ratios  
are [B/A]$_{R}$ = 0.94 $\pm$ 0.01, [C/A]$_{R}$ = 0.78 $\pm$ 0.01, 
[D/A]$_{R}$ = 0.74 $\pm$ 0.01. The R-band magnitudes
were taken from the CASTLES Web site.
The flux fraction of image A, [A/(A+B+C+D)], is 
a factor of 1.55 $\pm$ 0.17 larger in the full X-ray band than in the {\sl HST} R-band.  
  
The earlier spectral analyses of \clover\
used the combined spectrum of all images.
In our re-analysis of the data we search for possible spectral variations between images
by extracting a spectrum for image A and a combined spectrum of images
B, C, and D. The low S/N  of these data did not
allow us to extract individual spectra for images B, C, and D. 
The spectrum of image A was extracted 
from a circular region of radius 1\arcsec centered 0\sarc5 SW  from the center of image A
to minimize contamination from the other images.
The combined spectrum of images B, C, and D was extracted from a circular region centered 
on \clover\ with a radius of 2\sarc5, excluding the source region of  image A.
We determined corrections to the ancillary response files by simulating
the spectra of point sources at the locations of the images within the
apertures used in our analysis. 
For our simulations we used \verb+XSPEC+ to generate the source spectra and the ray-tracing tool \verb+MARX+ to  model the
dependence of photon scattering with energy.
The background was extracted from a source-free
annulus centered on \clover\ with inner and outer radii of 10\arcsec\ and
30\arcsec, respectively. All errors reported below are at the 90\%
confidence level unless noted otherwise. 
A variety of models were fitted to the spectrum of image
A and the combined spectrum of images B, C, and D,
employing the software tool \verb+XSPEC v11.2+ (Arnaud 1996). 
Spectral fits were performed in the 1--8~keV bandpass.
Given the relatively low-S/N of the spectra we performed fits
using the $C$-statistic  (Cash 1979) and compared the results with those obtained 
using the $\chi^{2}$ statistic. Spectral fits performed with the $C$-statistic
do not require binning of the data and produce reliable estimates of the 
best-fit parameters and their errors for low-S/N spectra. One of the limitations
of the $C$-statistic is that it cannot provide a goodness-of-fit measure.
However, simple analytic expressions do exist 
for estimating  the goodness-of-fit when using the  $\chi^{2}$ statistic.
We therefore used fits performed with the $\chi^{2}$ statistic
to evaluate the probabilities of exceeding the derived $\chi^{2}$ values.
For the spectral fits performed with the $\chi^{2}$ statistic we binned the data to
have a minimum of twelve counts per bin.
We did not find any significant differences
between the estimated best fit parameters and their associated errors
provided by the $C$-statistic and the $\chi^{2}$ statistic.
In Table 1 we list the best fit parameters and their errors based on fits using the $C$-statistic.

The spectra were initially fitted individually with a  
model consisting of a simple power law with
Galactic absorption due to neutral gas with a column density
of $N_{\rm H}$ = 1.8 $\times$ 10$^{20}$~cm$^{-2}$ (Stark et al. 1992).
The model also included neutral intrinsic absorption
at $z = 2.55$.
The fit residuals in image A showed an emission feature between 1.65--1.85~keV 
that led to an unacceptable fit, with a $\chi^{2}$ =  13.56 for 10  degrees of freedom (dof).
Only the spectrum of image A shows clear evidence of this strong emission
line. We modeled the residual feature in image A by considering a
Gaussian component near 1.7~keV in the observed frame. 
This led to a significant
improvement in fit quality at the $\sim$ 98\% confidence level (according to the $F$-test, $F=5.6$)
with a $\chi^{2}$ = 4.0 for 7 dof. 
In Figure 2a we show the spectrum of image A.
Protassov et al. (2002), have recently argued that the $F$-test cannot be applied when 
the null values of the additional parameters fall on the boundary of the allowable 
parameter space. They proposed a Monte Carlo approach to determine the distribution of the $F$ statistic. 
We followed this approach and
constructed the probability density distribution of the $F$-statistic between
spectral fits 1 and 2 of Table 1.  Our Monte Carlo simulations
indicate that the probability of obtaining an $F$ value
of 5.6 or greater is $\sim$1.5 $\times$ 10$^{-2}$ close to the result
found by using the analytic expression for the distribution of the $F$-statistic.
The best-fit rest-frame energy of the emission feature 
is $E_{line}$ = 6.3$\pm0.1$~keV.
Adding an unresolved Gaussian line to the combined spectrum of images B, C, and D leads to a slight
improvement in fit quality at the 86\% confidence level (according to the $F$-test).
The combined observed-frame spectrum of images B, C, and D fit with Galactic absorption, 
neutral absorption at the source, and a power-law model is shown in Figure 2b. 
The best-fit values for the photon indices in
the spectrum of image A and the combined spectrum of
images B, C, and D, are  $\Gamma$ = 1.8$\pm$0.8
and $\Gamma = 1.7\pm0.7$, respectively (see fits 1 and 4 of Table 1).
Our fits support the presence of an intrinsic absorber towards image A 
with a column density of $N_{\rm H}$ = (2.3$\pm$2.2) $\times$ 10$^{23}$~cm$^{-2}$ 
(see fit 1 of Table 1). We also find a significant intrinsic absorber from the fit to the 
combined spectrum of images B, C, and D of $N_{\rm H}$ = (3$_{-2}^{+3}$) $\times$ 10$^{23}$~cm$^{-2}$.
The observed column in images B, C, and D (see fit 4 of Table 1) may represent the
column density perpendicular to the outflow rather that along the outflow
as shown in Figure 3 of Oshima et al. (2001).
They estimated the  column along the flow to be $N_{H} \sim 10^{24}$~cm$^{-2}$. 
More complex models including ionized absorption or
partial covering were not considered in our analysis due to the low S/N of
the available spectrum and the high $z$ of \clover\
which results in several absorption edges due to Ne, O, N, and C 
being redshifted outside the \chandra\ bandpass.

We performed simultaneous fits to the spectrum
of image A and the combined spectrum of images B, C, and D to
estimate the significance of the difference in the
Fe~K$\alpha$ lines. The common spectral model for the fit consisted of
(a) Galactic absorption due to neutral gas with a column density
of $N_{\rm H}$ = 1.8 $\times$ 10$^{20}$~cm$^{-2}$,
(b) neutral absorption at the redshift of the quasar,
(c) a simple power law, and (d) a Gaussian component near 1.7~keV in the observed frame.
The photon index $\Gamma$ and the energies and widths of the Fe~K$\alpha$ lines were linked
as common parameters in this spectral fit. The best-fit parameters and 90\% confidence 
errors are presented in Table 1.
The rest-frame equivalent widths and 90\% confidence levels of the lines in the spectrum of image A
and the combined spectrum of images B, C, and D are 
$EW_{A}$ = 1.27$_{-0.59}^{+0.59}$~keV 
and $EW_{BCD}$ = 0.20$_{-0.20}^{+0.36}$~keV, respectively.
In Figure 3 we show the 68.3\% and 85\% confidence contours of the normalizations
of the Fe~K$\alpha$ lines versus the rest-frame Fe~K$\alpha$ line energy. 
We chose to show the 85\% contours instead of the commonly used 90\% contours
because the 85\% contours touch exactly in this case. Based on these contours, the probability that the iron lines in 
the spectrum of image A and the combined spectrum of images  B, C, and D have 
strengths that are consistent with each other is less than 0.02.
For this confidence contour plot we fixed the parameters of the power-law component of the model to their
best-fit values listed in fit 5 of Table 1. 
We conclude that the Fe~K$\alpha$ line in the spectrum
of image A is significantly enhanced by a factor of about six from that
of the combined spectrum of images B, C, and D.

\section{Discussion}

Our re-analysis of the X-ray spectra of the individual lensed images
of \clover\ indicates the presence of a microlensing event in image A
because (a) the flux fraction of image A, [A/(A+B+C+D)], is a factor of
1.55 $\pm$ 0.17 larger in the X-ray band than what was detected in the {\sl HST} R-band,
(b) a redshifted narrow iron line is present with a relatively high significance only in 
the spectrum of image A.  
To obtain additional insight on the possible energy dependence of the
magnification of image A, we estimated relative enhancements of image fluxes
in the continuum and iron-line emission components
based on fit 5 of Table 1.
Specifically, we find that the ratio of the X-ray fluxes of
image A and the combined images B, C and D
in the iron line component
is [A/(B+C+D)]$_{\rm Fe~{K\alpha}}$ $ >  $ 1.7 at the 85\% confidence level (see Figure 2).
The ratio of the X-ray flux densities at 1~keV (observed frame)
in the continuum component
is [A/(B+C+D)]$_{\rm Cont}$ = 0.8 $\pm$ 0.2 (90\% confidence intervals).
These values differ significantly from the {\sl HST} R-band flux ratio
of [A/(B+C+D)]$_{\rm R}$=0.41$\pm$0.01
which is expected to be less affected by a microlensing event
produced by a star or group of stars in the foreground lensing galaxy since 
the characteristic sizes of the X-ray 
continuum emission regions of quasars are expected 
to be considerably less than the sizes of the optical R-band emission regions. 
Specifically, variability studies indicate that the sizes of the X-ray and optical continuum emission regions
are  $\sim$ 4$\times 10^{15-16}$~cm  and
$\sim$ 4$\times 10^{16-17}$~cm, respectively (e.g., Chartas et al. 2001; Dai et al 2002; Wyithe et al. 2000).
We note that since the flux fraction of image A in the X-ray
continuum is greater than the flux fraction in the R band, the 
large equivalent width of the Fe line in image A
is not caused by a decrease in the X-ray continuum but is more likely produced 
by an amplification of the flux of the line.

A scenario in which intrinsic variability is the cause of the detection
of a strong iron line in only image A and a significantly weaker line in the 
combined spectrum of images B, C and D 
requires fine tuning such that the flare of the iron line
needs to occur within a certain window in time for it to be detected with a 
large $EW$ in only one image and the flare must 
last for only a few days. Our simple two singular isothermal sphere lens model for \clover\ predicts
that image B leads images C, A, and D by 5~days, 9~days and 16~days respectively.
A similar lens model by Chae \& Turnshek (1999) was found to reproduce the observed relative 
amplifications fairly accurately.
Another strong argument against the intrinsic variability interpretation is that changes of
factors of $\sim$ 6 in the equivalent width of a fluorescent Fe~K${\alpha}$ line in a
Seyfert 1 galaxy over periods of a few days have not been detected to date
(e.g., Markowitz, Edelson \& Vaughan 2003; Gliozzi, Sambruna \& Eracleous 2003; Lee et al. 2000; 
Leighly 1999).
We anticipate that the amplitude of Fe~K${\alpha}$ line variability to be less in quasars than in 
Seyfert 1 galaxies due to the larger sizes and luminosities of quasars.
X-ray variability, at least of the continuum emission in Seyfert 1 galaxies and quasars,
is found to be anticorrelated with X-ray luminosity
(e.g., Nandra et al. 1997; Turner et al. 1999; George et al. 2000; Dai et al. 2003).
A more likely explanation of the observed energy dependent magnification in the X-ray band 
and the enhancement of image A in the X-ray band compared to that in the optical is a 
microlensing event in image A. 
Such an event will not lead to time-delayed magnifications in the remaining images and therefore 
can explain the non-detection of the Fe~K${\alpha}$
line in the remaining images. A characteristic size scale commonly used in 
microlensing analysis is the projected 
Einstein ring radius of the star, $\zeta_{E}$. Emission regions with a size significantly
larger than $\zeta_{E}$ will not be affected by microlensing,
whereas, emission regions less than $\zeta_{E}$ will be significantly magnified.
For \clover\ with lens and source redshifts of 1.7 and 2.55 
respectively, and assuming an isolated star, the Einstein-ring radius
on the source plane is 
$\zeta_{E}$ $\sim$ 2 $\times$ 10$ ^{16}$$(M_{\rm star}/M_{\odot})^{1/2}$$h^{-0.5}_{75}$ cm.

The observed enhancement of the Fe~K${\alpha}$ line implies that the
size of this iron line region is $R_{line}$ $ \simlt $ $\zeta_{E}$.
The enhanced Fe~K$\alpha$ emission observed in image A may be produced by 
microlensing of Fe~K$\alpha$ fluorescence from the far side of the outflow. 
Assuming the Murray et al. (1995) disk-wind geometry for BALQSOs
we expect that Fe~K$\alpha$ fluorescence from the near-neutral accretion disk, 
thought to be emitted (based on modeling of the Fe~K$\alpha$ line profiles in several Seyferts)
from annuli of inner radii of $r_{in}$ $\sim$ 6~$r_{g}$
and outer radii of $r_{out}$ $\sim$ 30~$r_{g}$ (e.g., Nandra et al. 1997),
where $r_{g} = GM_{bh}/c^{2}$ and $M_{bh}$ is the mass of the black hole of \clover,
will be heavily absorbed by the quasar outflow.
The far side of the wind is a more plausible source of the Fe~K$\alpha$ line because our line of sight to it does not traverse a 
significant absorbing column.
We estimate the size of the microlensed fluorescent region of the quasar outflow 
to be  $\simlt$ $\zeta_{E}$. We have assumed a minimum launching radius for the UV BAL absorber
of $r_{\rm min} \sim 1 \times 10^{16}(M_{bh}/10^{8} M_{\odot})^{1/2}$~cm 
based on hydrodynamical simulations performed by Proga et al. (2000).
To estimate $M_{bh}$ we assumed the lensed bolometric luminosity of
$L_{Bol}$ = 3.9 $\times$ 10$^{47}$ $\mu^{-1}$~erg~s$^{-1}$ (Granato et al. 1996), 
$\mu$ is the magnification factor ranging between 10 and 23 based
on lens models of this system  (e.g., Venturini \& Solomon 2003, Chae et al. 1999), $L_{Bol}/L_{Edd}$ = $\eta$, 
$L_{Edd}$ is the Eddington luminosity and $\eta$ is thought to range between 0.1 and 1 
in quasars. We find, $L_{Edd}$ = 3.9 $\times$ 10$^{47}$ $\mu^{-1}\eta^{-1}$~erg~s$^{-1}$, 
$r_{g}$ =  4.5 $\times$ 10$^{14}$~$\mu^{-1}\eta^{-1}$~cm,
$M_{bh}$ = 3 ~$\times$~10$^{9}$~$\mu^{-1}\eta^{-1}$~$M_{\odot}$
and $r_{\rm min}$= 5.5 $\times$ 10$^{16}$~$\mu^{-1/2}\eta^{-1/2}$~cm.
We note that with the values we 
have adopted the product $\mu\eta$ is approximately equal to 1. 
The relatively smaller magnification of the observed X-ray continuum emission 
compared to the line emission
suggests that the size of the region producing the observed X-ray continuum emission 
is larger than $\zeta_{E}$.

The processes thought to contribute to 
the continuum X-ray emission in radio-quiet quasars
include inverse Compton scattering of photons in disk coronae by UV photons 
originating in the accretion disk, resulting in an increase in the energy 
of the upscattered photons from the UV range to the 2-10~keV range.
At rest energies above $\sim$ 10~keV, Compton scattering of photons from the
disk corona by the disk becomes significant.
Variability studies of radio-quiet quasars with similar black
hole masses to \clover\ indicate that the size of the emission region responsible
for these continuum processes described above is of order 10$^{15-16}$~cm (e.g., Chartas et al. 2001; Dai et al 2002.)
This, combined with our conclusion that the observed X-ray continuum emission
must have a size  $ > > $ $\zeta_{E}$ suggests that the X-ray continuum 
emission in \clover\ is not viewed directly
but most likely is reflected in a scattering
region with a characteristic size of $r_{scat}$ $ > > $ $\zeta_{E}$.

Our constraint on the size of the scattering region is supported by
recent {\sl HST} observations of \clover;
using imaging polarimetry with the {\sl HST}, Chae et al. (2001) detected a
0.07 mag dimming and a change of the linear polarization
component of image D relative to the other images
during the June 1999 observation which they interpreted as microlensing of a
portion of a large-scale, asymmetric continuum scattering region seen in polarized light.
They also conclude that the microlensing caustic
during their observations was located away from the continuum emitting region.
They independently conclude that the scattering region is
10$^{18}L^{0.5}_{46}$ cm $ >  $ $r_{scat}$  $ > $ 2 $\times$ 10$ ^{16}$$(M_{\rm star}/M_{\odot})^{1/2}$$h^{-0.5}_{75}$ cm. 
We also note that based on the simulations of 
Witt, Mao, and Schechter (1995) images A and D of \clover\ have the 
largest probability of being microlensed due to the relatively large
optical depths for microlensing events along these lines of sight.
In Figure 4 we present a sketch of a scenario  that can explain the observed X-ray spectra of \clover.
For simplicity the caustic plane is oriented perpendicular to the observer's line of sight. 
In this proposed scenario the microlensing caustic is centered near the black hole
and the observed continuum emission in all images originates
predominatly from a large scale scattering region.
We expect the continuum X-ray emission region
and the Fe~K$\alpha$ fluorescence from the near-neutral accretion disk
to be significantly absorbed along our line of sight.
This is supported by the fact that \clover\ is a LoBAL quasar
and recent studies indicate that their X-ray spectra may be reflection/scattered dominated
(e.g., Green et al., 2001; Gallagher 2002).
Based on the anticorrelation between $\alpha_{ox}$ and 2500~${\rm \AA}$ 
luminosity (Vignali et al. 2003) for radio-quiet quasars (RQQs) we estimate an
average value of $\alpha_{ox}$  $\sim$ -1.6 $\pm$ 0.2 for a RQQ with 
$l_{2500\rm \AA}$ = 6.1 $\times$ 10$^{30}$~erg~s$^{-1}$~Hz$^{-1}$ of \clover.
In a recent survey of high redshift lensed quasars
Dai et al. (2003) calculated 
$\alpha_{ox}$($\alpha_{ox}$ corrected) = $-3.3 \pm 0.5(-1.8_{-0.1}^{+0.15})$
for H~1413+117  based on the X-ray flux of the images B, C, and D
(image A was not included since it is likely microlensed) 
confirming the X-ray weakness of this LoBAL.  The estimated values of 
$\alpha_{ox}$ are highly dependent on the assumed model for the intrinsic absorption.
The observed X-ray emission in image A also contains
significantly magnified Fe~K$\alpha$ emission originating from the far side of the flow
and less magnified emission from a portion of the large scale scattering region.

Another possiblity is that the microlensing caustic during the {\it Chandra} observations
was not centered on the black hole but located further away than the X-ray continuum emission region
of the inner accretion disk.
In this scenario the observed continuum emission in all images originates
from both the continuum-emitting region near the inner  accretion disk
and the large-scale scattering region.
The X-ray flux observed in image A also contains
magnified emission from the far side of the outflow and magnified emission
from a portion of the large scale scattering region.
Since the microlensing caustic in this case does not overlap with the central region of the black hole
complete obscuration of the emission from the inner region of the accretion disk
is not required to explain the observed energy-dependent magnification.
To summarize, the various emission regions
referred to in this analysis in order of smallest to largest regions are:
(a) A very small-scale ($ < < \zeta_{E}$), possibly obscured
Fe~K$\alpha$ fluorescence region. (b) A small-scale ($ \simlt \zeta_{E}$) portion of the
reflection region from the far side of the outflow,
which is relatively cold and
neutral, slightly redshifted, and is highly amplified by microlensing.
This region causes the large equivalent width for the iron line in image A.
(c) A small-scale ($\simlt  \zeta_{E}$) X-ray continuum emission region
originating near the inner-accretion disk.
(d) A mid-scale ($\sim \zeta_{E}$) optical continuum R band emission region, and,
(e) A large-scale ($ > \zeta_{E}$)  X-ray and optical scattering region which is highly ionized and
reflects the X-ray source without creating a fluorescent iron line.

Optical spectra of \clover\ taken with the SILFID spectrograph at the 
Canada-France-Hawaii ({\sl CFH}) 
telescope in 1989 (Angonin et al. 1990)
and the {\sl HST} Space Telescope Imaging Spectrograph in 2000 (spectra taken by E. M. Monier
cited in Chae et al. 2001) indicate that a microlensing event
along the sight-line of image D was in progress for 
at least a duration of 11 years. This timescale is consistent with 
the estimate by Witt, Mao, \& Schechter (1995)
of the time for the source to cross an Einstein radius for \clover.  
We may, therefore, expect that the possible microlensing event
in image A persists for a similar timescale.

The detection of the iron line is possible perhaps 
because of an ongoing microlensing event in image A.
Future observations of \clover\ with \chandra\ may provide a confirmation
of the microlensing interpretation if a significant variation in the 
continuum and/or iron line flux ratios in the images is detected. 
This confirmation would validate
our microlensing models and our conclusions regarding the relative
size of the continuum and iron line scatterers.

\acknowledgments
We acknowledge financial support from NASA grants NAG5-9949, NAS 8-38252
and NAS 8-01128.  We thank the anonymous referee for helpful comments and suggestions. \\

\small

\clearpage

\normalsize

\beginrefer
\noindent
\refer Angonin, M.-C., Vanderriest, C., Remy, M., \& Surdej, J.\ 1990, \aap, 233, L5 \\

\refer Arnaud, K.~A.\ 1996, ASP 
Conf.~Ser.~101: Astronomical Data Analysis Software and Systems V, 5, 17 \\

\refer Barvainis, R., Antonucci, R., Hurt, T., Coleman, P., \& Reuter, H.-P.\ 1995, \apjl, 451, L9\\

\refer Boroson, T.~A.~\& Meyers, K.~A.\ 1992, \apj, 397, 442 \\

\refer Cash, W.\ 1979, \apj, 228, 939 \\

\refer Chartas G., 2000, \apj, 531, 81. \\

\refer Chartas, G., Dai, X., Gallagher, S.~C., Garmire, G.~P., Bautz, M.~W., Schechter, P.~L., \& Morgan, N.~D.\ 2001, \apj, 558, 119 \\

\refer Chartas, G., Brandt, W.~N., Gallagher, S.~C., \& Garmire, G.~P. \ 2002,
\apj, 579, 169\\

\refer Chartas, G., Brandt, W.~N., Gallagher, S.~C., \& Garmire, G.~P. \ 2003,
Astron. Nachr., 324, 173\\

\refer Chae \& Turnshek\ 1999, ApJ, 514 587 \\

\refer Chae, K., Turnshek, D.~A., Schulte-Ladbeck, R.~E., Rao, S.~M., \& Lupie, O.~L.\ 2001, \apj, 561, 653 \\

\refer Dai, X., Chartas, G., Agol, E., Bautz, M.~W., \& Garmire, G.~P.\ 2003, \apj, 589, 100 \\

\refer Dai, X., Chartas, G., Eracleous, M., \& Garmire, G.~P.\ 2003, submitted to \apj \\

\refer Gallagher, S.~C., Brandt, W.~N., Chartas, G., \& Garmire, G.~P.\ 2002,  \apj, 567, 37 \\

\refer George, I.~M., Turner, T.~J., Yaqoob, T., Netzer, H., Laor, A., Mushotzky, R.~F., Nandra, K., \& Takahashi, T.\ 2000, \apj, 531, 52 \\

\refer Gliozzi, M., Sambruna, R.~M., \& Eracleous, M.\ 2003, \apj, 584, 176\\

\refer Granato, G.~L., Danese, L., \& Franceschini, A.\ 1996, \apjl, 460, L11 \\

\refer Green, P.~J., Aldcroft,  T.~L., Mathur, S., Wilkes, B.~J., \& Elvis, M.\ 2001, \apj, 558, 109 \\

\refer Hasinger, G., Schartel, N., \& Komossa, S.\ 2002, \apjl, 573, L77 \\

\refer Hazard, C., Morton, D.~C., Terlevich, R., \& McMahon, R.\ 1984, \apj, 282, 33\\

\refer Kneib, J.-P., Alloin, D., \& Pello, R.\ 1998, \aap, 339, L65 \\

\refer Lee, J.~C., Fabian, A.~C., Reynolds, C.~S., Brandt, W.~N., \& Iwasawa, K.\ 2000, \mnras, 318, 857\\

\refer Leighly, K. M. 1999, ApJ, 125, S317\\

\refer Magain, P., Surdej, J., Swings, J.-P., Borgeest, U., \& Kayser, R.\ 1988, \nat, 334, 325 \\

\refer Markowitz, A., Edelson, R., \& Vaughan, S.\ 2003, \apj, in press \\

\refer Monier, E.~M., Turnshek, D.~A., \& Lupie, O.~L.\ 1998, \apj, 496, 177 \\

\refer Mori, K., Tsunemi, H., Miyata, E., Baluta, C., Burrows, D. N.,
Garmire, G. P., \& Chartas, G. 2001, in ASP Conf. Ser. 251, New Century
of X-Ray Astronomy, ed. H. Inoue \& H. Kunieda (San Francisco: ASP), 576 \\

\refer Nandra, K., George, I.~M., Mushotzky, R.~F., Turner, T.~J., \& Yaqoob, T.\ 1997, \apj, 476, 70 \\

\refer Nandra, K., George, I.~M., Mushotzky, R.~F., Turner, T.~J., \& Yaqoob, T.\ 1997, \apj, 477, 602 \\

\refer Oshima, T., Mitsuda, K., Fujimoto, R., Iyomoto, N., Futamoto, K., Hattori, M., Ota, N.,
Mori, K., Ikebe, Y., Miralles, J.~M., \&  Kneib, J.-P. \ 2001, \apjl, 563, L103 \\

\refer Protassov, R., van Dyk, D.~A., Connors, A., Kashyap, V.~L., \& Siemiginowska, A.\ 2002, \apj, 571, 545 \\

\refer Shalyapin, V.~N.\ 2001, Astronomy Letters, 27, 150 \\

\refer Stark, A.~A., Gammie, 
C.~F., Wilson, R.~W., Bally, J., Linke, R.~A., Heiles, C., \& Hurwitz, M.\ 
1992, \apjs, 79, 77 \\

\refer Tsunemi, H., Mori, K., 
Miyata, E., Baluta, C., Burrows, D.~N., Garmire, G.~P., \& Chartas, G.\ 
2001, \apj, 554, 496 \\

\refer Turner, T.~J., George, I.~M., Nandra, K., \& Turcan, D.\ 1999, \apj, 524, 667 \\

\refer Vignali, C., Brandt, W.~N., Schneider, D.~P., Anderson, S.~F., Fan, X., 
Gunn, J.~E., Kaspi, S., Richards, G.~T., \& Strauss, M.~A., 2003, \aj, 125, 2876 \\

\refer Wise, M. W., Davis, J. E., Huenemoerder, Houck, J. C., Dewey, D.
Flanagan, K. A., and Baluta, C. 1997,
{\it The MARX 3.0 User Guide, CXC Internal Document}
available at http://space.mit.edu/ASC/MARX/ \\

\refer Witt, H.~J., Mao, S., \& Schechter, P.~L.\ 1995, ApJ, 443, 18 \\

\refer Wyithe, J.~S.~B., Webster, R.~L., Turner, E.~L., \& Mortlock, D.~J.\ 2000, \mnras, 315, 62 \\

\refer Yonehara, A.\ 2001, \apjl, 548, L127 \\

\endrefer

\newpage
\scriptsize
\begin{center}
\begin{tabular}{cclcc}
\multicolumn{5}{c}{TABLE 1}\\
\multicolumn{5}{c}{RESULTS FROM FITS TO THE SPECTRA OF H~1413+117} \\
 & & &  &\\ \hline\hline
\multicolumn{1}{c} {Fit} &
\multicolumn{1}{c} {Source$^{1}$} &
\multicolumn{1}{c} {Model} &
\multicolumn{1}{c} {Parameter$^{2}$} &
\multicolumn{1}{c} {Value$^{3}$} \\
 & &                               &                                                                   \\
1&H~1413+117$^{4}$ & PL, neutral absorption &   $\Gamma_{A}$           & 1.8$_{-0.8}^{+0.8}$                 \\
&(Image A)  &  at source, and a Gaussian  &   $N_{\rm H,A}$          & (2.3$\pm2.2$) $\times$ 10$^{23}$~cm$^{-2}$        \\
&  &  emission line at source.            &   E$_{line}$             & 6.3$_{-0.1}^{+0.1}$~keV                 \\
&  &  The fit is performed to the         &   $\sigma_{line}$        & $ < $ 0.14~keV                        \\
&  &  spectrum of image A of H1413.       &  EW$_{line}$             & 1.1$_{-0.5}^{+0.8}$~keV                 \\
&  &                                      & C-statistic/{nbins}$^{5}$& 4.5/13                                         \\
&  &                                      & $\chi^2/{\nu}$           & 4.0/7                                         \\
&  &                                      & $P(\chi^2/{\nu})^{6}$    & 0.78                                  \\
&  &                                      &                          &                                          \\
2&H~1413+117  & PL, neutral absorption      & $\Gamma_{A}$             &  2.7$_{-0.7}^{+0.7}$                 \\
&(Image A)  &  at source.                 & $N_{\rm H,A}$            & (5$\pm2$) $\times$ 10$^{23}$~cm$^{-2}$                                          \\
&  &  The fit is performed to             & C-statistic/{nbins}      & 19.2/13                               \\
 & &  the spectrum  of image A of H1413.  & $\chi^2/{\nu}$           & 13.6/10                            \\
&  &                                      & $P(\chi^2/{\nu})$        & 0.19                                 \\
&  &                                      &                          &                                      \\
3&H~1413+117 & PL, neutral absorption       & $\Gamma_{BCD}$           &  1.7 {f}$^{7}$                  \\
&(Images B + C + D ) & at source, and a narrow Gaussian              & $N_{\rm H,BCD}$    & (3$_{-2}^{+3}$) $\times$ 10$^{23}$~cm$^{-2}$        \\
&  &  emission line at source.            &   E$_{line}$             & 6.4$_{-0.1}^{+0.1}$~keV                 \\
&  &  The fit is performed to the         &  $\sigma_{line}$         &  0.01~keV {f}                        \\
&  &  the combined spectrum               &  EW$_{line}$             & 0.35$_{-0.35}^{+0.50}$~keV                                         \\
&  &  of images B, C, and D.              & C-statistic/{nbins}& 10.64/12                                         \\
&  &                                      & $\chi^2/{\nu}$           & 7.42/8                            \\
&  &                                      & $P(\chi^2/{\nu})$        & 0.49                                 \\
&  &                                      &                          &                                          \\
4&H~1413+117  & PL, neutral absorption      &$\Gamma_{BCD}$            &  1.7$_{-0.7}^{+0.7}$                 \\
&(Images B + C + D)  &  at source.        & $N_{\rm H,BCD}$          & (4$_{-2}^{+3}$) $\times$ 10$^{23}$~cm$^{-2}$                               \\
 & &  The fit is performed to             & C-statistic/{nbins}    & 10.64/12                                           \\
  &&  the combined spectrum               & $\chi^2/{\nu}$         & 9.87/9                                  \\
  &&  of images B, C, and D.              & $P(\chi^2/{\nu})$      & 0.36                                          \\
  &&                           &                   &                                          \\
&  &                                      &                          &                                          \\
5&H~1413+117  & PL, neutral absorption      &$\Gamma$            &  1.6$_{-0.9}^{+0.8}$                 \\
& Simultaneous fit  &  at source and a narrow Gaussian        & $N_{\rm H}$        & (2.4$_{-2.4}^{+2.8}$) $\times$ 10$^{23}$~cm$^{-2}$                       \\
& to images A and  &  emission line at source.            &   E$_{line}$             & 6.3$_{-0.1}^{+0.1}$~keV                 \\
& B+C+D &  The fit is performed          &  $\sigma_{line}$         &  0.06$_{-0.06}^{+0.3}$~keV                         \\
&  &   simultaneously to the              &  EW$_{A}$             & 1.27$_{-0.59}^{+0.59}$~keV                       \\
&  &  the combined spectrum               &  EW$_{BCD}$             & 0.20$_{-0.20}^{+0.36}$~keV                        \\
&  &  of image A and            & C-statistic/{nbins}& 22.5/27                                         \\
  &&  the combined spectrum              & $\chi^2/{\nu}$         & 18.0/27                                  \\
  &&  of images B, C, and D.              & $P(\chi^2/{\nu})$      & 0.52                                          \\
  &&                           &                   &                                          \\
\hline \hline
\end{tabular}
\end{center}
${}^{1}$All model fits include Galactic absorption towards the source (Stark et al. 1992).\\
${}^{2}$All absorption-line parameters are calculated in the quasar rest frame.\\
${}^{3}$All errors are for 90\% confidence with all
parameters taken to be of interest except absolute normalization.
All best fit parameters and associated errors are taken from fits using the C-statistic.\\
${}^{4}$For results on the combined
spectrum of all images of H~1413+117 see also Oshima et al. (2000) and Gallagher et al. (2002).\\
${}^{5}$ nbins is the number of pulse height analysis (PHA) bins in the spectrum. \\
${}^{6}$ $P(\chi^2/{\nu})$ is the probability of exceeding $\chi^{2}$ for ${\nu}$ degrees of freedom.\\
${}^{7}$ The symbol f indicates that the parameter is frozen in the spectral fit.\\

\newpage
\begin{figure*}
\centerline{\includegraphics[width=16cm,angle=0]{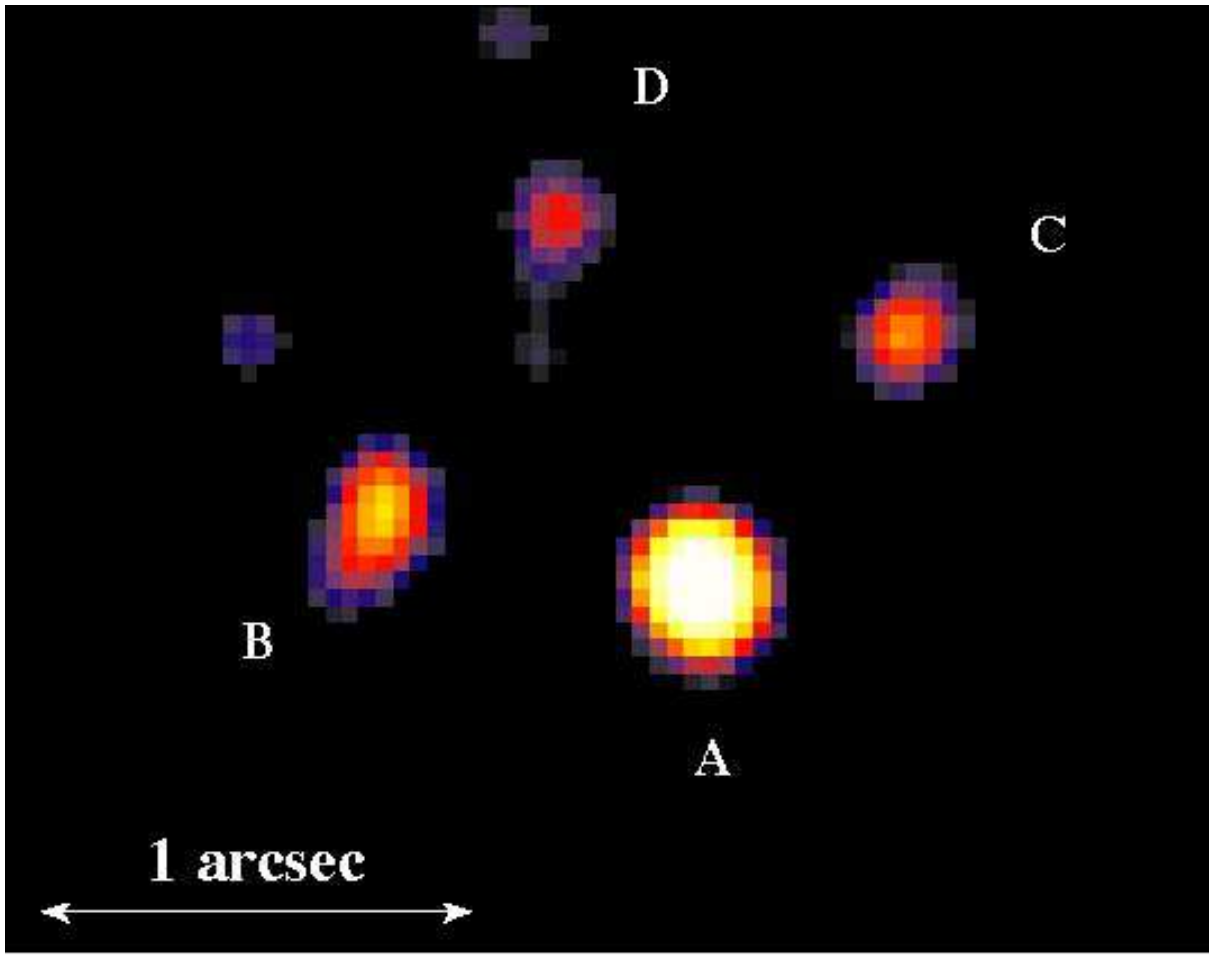}}
\caption{The deconvolved X-ray image of the Cloverleaf in the 1--8~keV bandpass.  
To improve the spatial resolution we employed a sub-pixel resolution technique
developed by Tsunemi et al. (2001) and later improved by Mori et al. (2001).
The X-ray flux fraction of image A, [A/(A+B+C+D)], is larger by a factor of 1.55 $\pm$ 0.17
than the optical R-band flux fraction, indicating that A is significantly enhanced in the X-ray band.
\label{fig:image}}
\end{figure*}

\newpage
\begin{figure*}
\centerline{\includegraphics[width=12cm,angle=0]{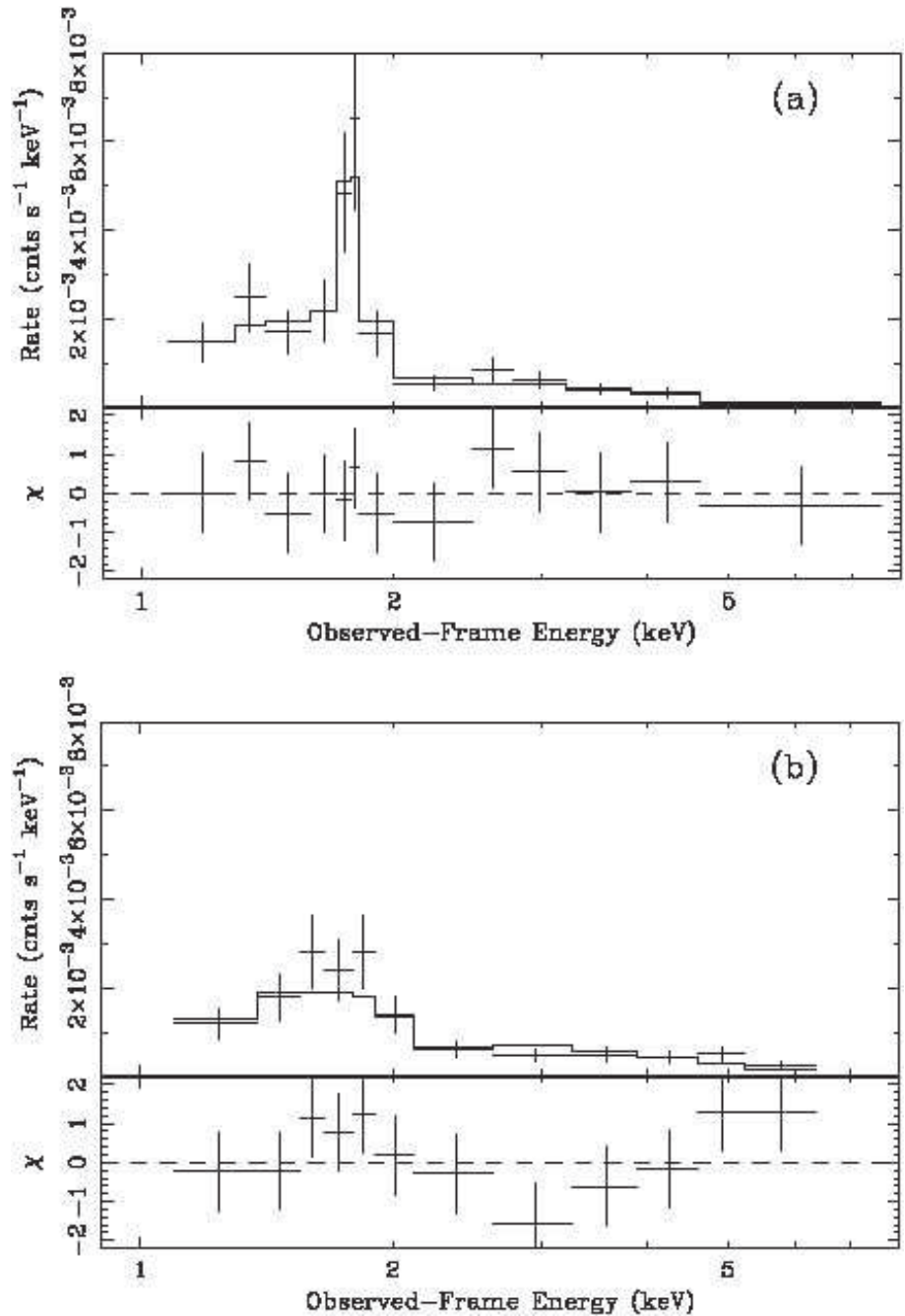}}
\caption{\small (a) The top panel shows the \chandra\ observed-frame spectrum of image A of  
\clover\ fit with Galactic absorption, neutral absorption at the source, a power-law continuum and 
a Gaussian emission line. In the lower panel the $\chi^{2}$ residuals of fit 1 of Table 1 indicate 
that this model can account for the spectral features in \clover.
(b) The top panel shows the \chandra\ observed-frame spectrum of the combined images B, C, and D of 
\clover\ fit with Galactic absorption, neutral absorption at the source, and a power-law continuum. 
In the lower panel we show the $\chi^{2}$ residuals of fit 4 of Table 1. 
The spectra have been binned for illustration purposes only. The significance of the Fe~K$\alpha$ 
line in each spectrum was inferred from fits
using the $C$-statistic that does not require binning of the data (see \S 2. for more details). 
 \label{fig:spec12}}
\end{figure*}

\newpage
\begin{figure*}
\centerline{\includegraphics[width=12cm,angle=-90]{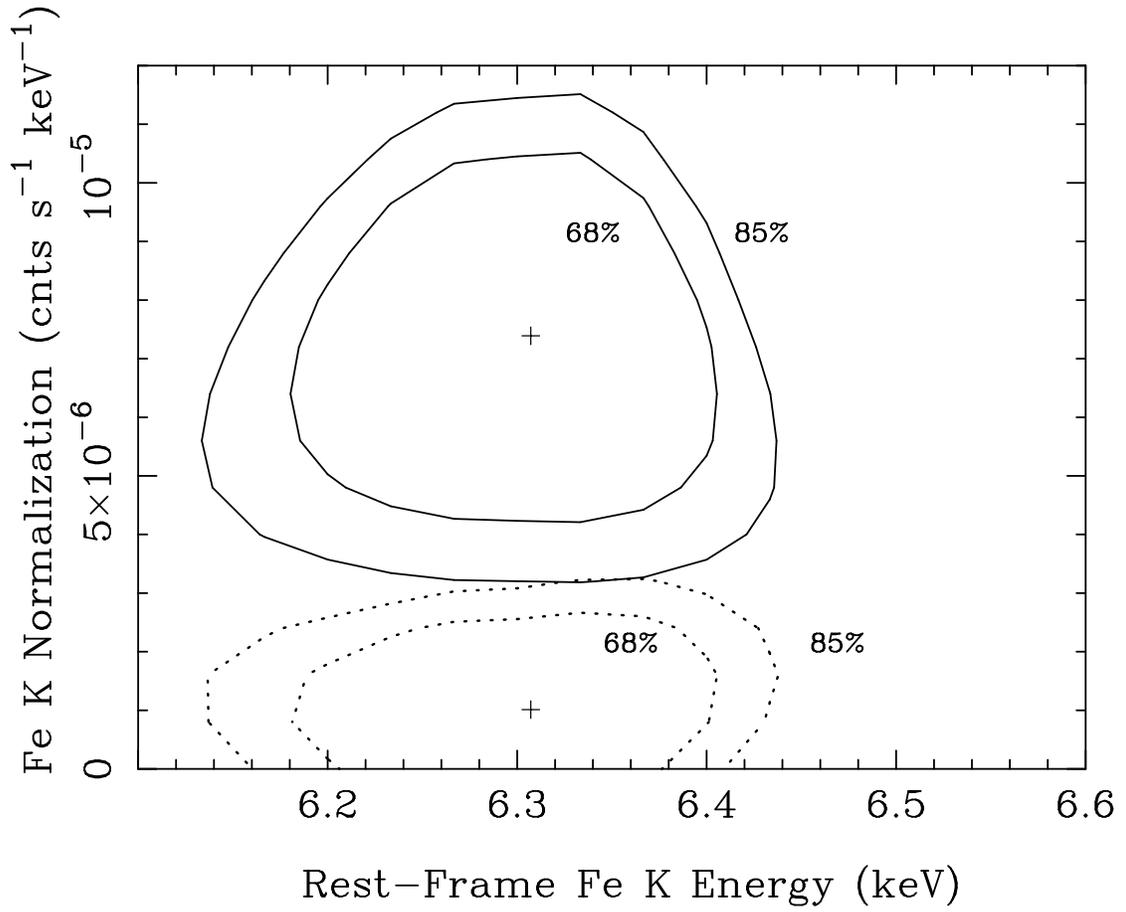}}
\caption{68.3\% and 85\% confidence contours of the normalizations
of the Fe~K$\alpha$ lines in the spectrum of image A (solid contours) and 
the combined spectrum of images B, C, and D (dotted contours)  versus the rest-frame Fe~K$\alpha$ line energy.
The parameters of the power-law component of the model were fixed to their
best-fit values listed in fit 5 of Table 1. 
\label{fig:image}}
\end{figure*}

\newpage
\begin{figure*}
\centerline{\includegraphics[width=16cm,angle=0]{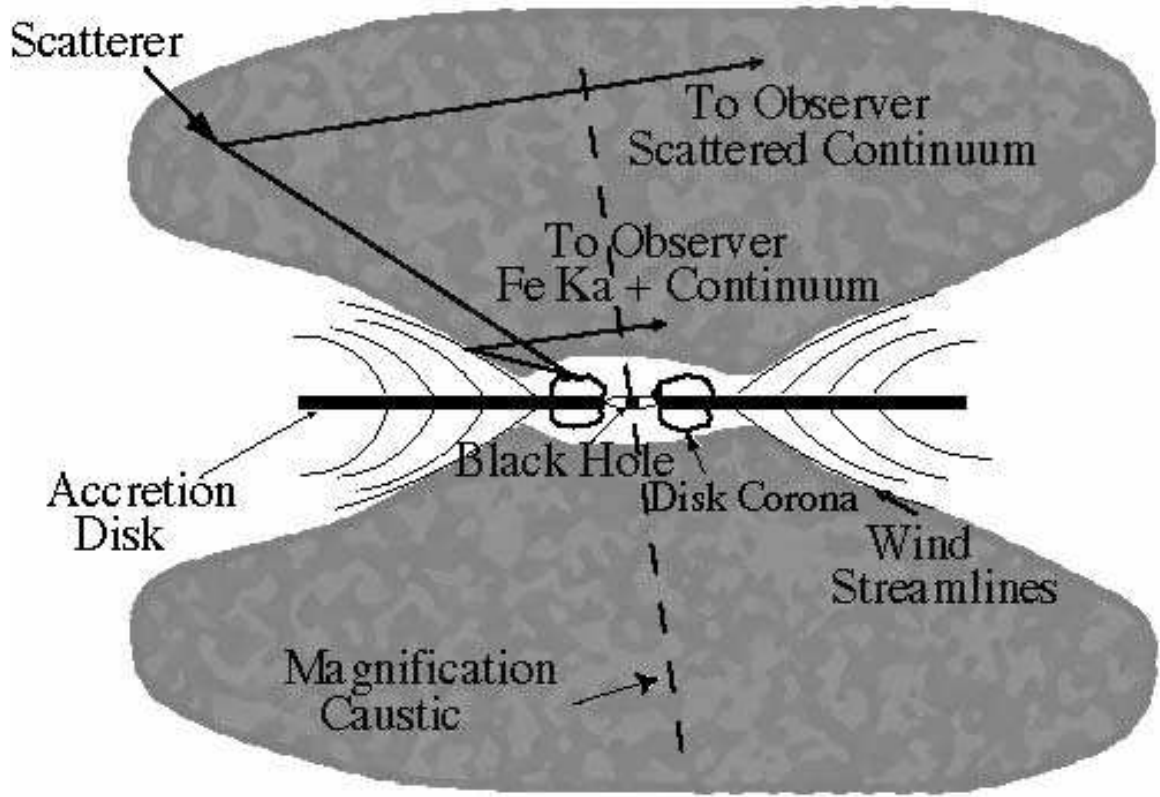}}
\caption{\small A proposed representation of a BALQSO disk wind overlayed with a microlensing
caustic (dashed line). This model represents the first scenario in the discussion section.
The caustic plane is oriented perpendicular to the observer's line of sight. 
In our proposed model Fe~K$\alpha$ emission produced from reprocessing off 
the BAL wind is located within the magnification caustic and suffers a 
significant magnification. Continuum X-ray emission produced from 
scattering off gas-clouds and the BAL wind is extended beyond the diamond-shaped caustic and suffers less magnification. 
 \label{fig:model}}
\end{figure*}

\end{document}